\documentclass{article}
\usepackage[T1]{fontenc}
\usepackage[utf8]{inputenc}
\usepackage{ismir} 
\usepackage{amsmath,amsfonts,cite,url,multirow}
\usepackage{graphicx}
\usepackage{color}
\usepackage{cleveref}

\title{Improving Neural Pitch Estimation with SWIPE Kernels}

\oneauthor
  {David Marttila, Joshua D. Reiss}
  {Centre for Digital Music, Queen Mary University of London\\\texttt{d.sudholt@qmul.ac.uk, joshua.reiss@qmul.ac.uk}}

\def\authorname{D. Marttila and J. D. Reiss}

\begin{document}

\maketitle

\begin{abstract}
Neural networks have become the dominant technique for accurate pitch and periodicity estimation. Although a lot of research has gone into improving network architectures and training paradigms, most approaches operate directly on the raw audio waveform or on general-purpose time-frequency representations. We investigate the use of Sawtooth-Inspired Pitch Estimation (SWIPE) kernels as an audio frontend and find that these hand-crafted, task-specific features can make neural pitch estimators more accurate, robust to noise, and more parameter-efficient. We evaluate supervised and self-supervised state-of-the-art architectures on common datasets and show that the SWIPE audio frontend allows for reducing the network size by an order of magnitude without performance degradation. Additionally, we show that the SWIPE algorithm on its own is much more accurate than commonly reported, outperforming state-of-the-art self-supervised neural pitch estimators.
\end{abstract}

\section{Introduction}

Pitch plays a central role in how humans perceive sound. Consequently, pitch estimation is a fundamental task in many music, speech and audio processing pipelines. While pitch is a psychoacoustic phenomenon, it closely correlates to the signal processing concept of the fundamental frequency $f_0$. Recent literature commonly uses the term ``pitch estimation'' to refer to the task of estimating an audio signal's $f_0$.

Given the importance of accurate pitch estimation, the topic has received a considerable amount of research attention over the past decades. Numerous digital signal processing (DSP) techniques estimate pitch based on the cepstrum \cite{nollCepstrumPitchDetermination1967}, the power spectrum \cite{maher1994fundamental, camachoSawtoothWaveformInspired2008, gonzalezPEFACPitchEstimation2014}, or the autocorrelation function \cite{DeCheveigne2002, mcleodSmarterWayFind2005a}.

More recently, deep neural networks have been applied to the task of pitch estimation \cite{kimCREPEConvolutionalRepresentation2018, ardaillonFullyConvolutionalNetworkPitch2019, morrisonCrossdomainNeuralPitch2023a, liYOLOPitchTimeFrequencyDualBranch2024}. In a typical architecture, a convolutional neural network (CNN) is given overlapping frames of raw audio as input and trained to predict a probability distribution over a discrete set of $f_0$ candidates in a supervised fashion. While these models can reach very high accuracy, they require a large amount of training data annotated with reliable ground truth pitch values and can struggle with out-of-domain generalization and robustness to noise and reverberation. Additionally, the CNNs usually consist of millions of parameters, making them less suitable for use in low-resource environments.

These drawbacks have been addressed in two different ways. Self-supervised training paradigms \cite{9053798, gfellerSPICESelfSupervisedPitch2020,  engelSelfSupervisedPitchDetection2020, riouPESTOPitchEstimation2023} do not require labeled training data, and incorporating traditional DSP approaches into neural networks has been shown to increase efficiency and robustness \cite{rengaswamy$$hf_0$$HybridPitch2021, hassanHAEPFHybridApproach2024}.

In this paper, we combine these two approaches and merge task-specific DSP-based features with both supervised and self-supervised training paradigms. Specifically, we substitute the audio frontend in the \textit{Pitch Estimation with Self-Supervised Transposition-Equivariant Objective} (PESTO) architecture \cite{riouPESTOPitchEstimation2023} for a representation obtained from the \textit{Sawtooth Waveform Inspired Pitch Estimator} (SWIPE) \cite{camachoSawtoothWaveformInspired2008}, which estimates pitch by measuring the similarity of the input spectrum to that of sawtooth waves at various pitch candidates. We also investigate the use of SWIPE as a frontend for supervised neural pitch estimators, which usually operate directly on the audio waveform.

The core insights of our work are these:

\begin{itemize}
    \item Although SWIPE is commonly used as a baseline for neural pitch estimation, we find that its performance has been significantly underreported. We show that SWIPE in its original form surpasses the accuracy of the state of the art in self-supervised neural pitch detection (PESTO).
    
    \item SWIPE is a well-suited audio frontend for neural pitch estimators in both supervised and self-supervised settings, and can improve the state-of-the-art in terms of accuracy, robustness, efficiency, and latency.
\end{itemize}

Trained models alongside a SWIPE implementation in PyTorch \cite{NEURIPS2019_9015} are available online.\footnote{\url{https://github.com/dsuedholt/neural-pitch-swipe}} The remainder of this paper is structured as follows. In Section~\ref{sec:background}, we give an overview over SWIPE and neural pitch estimation methods. Section~\ref{sec:methods} covers some aspects of our SWIPE implementation choices and details how we embed SWIPE into neural pitch estimation architectures. Section~\ref{sec:setup} describes how we evaluate our approach, and Section~\ref{sec:results} presents the results of our evaluation.

\section{Background}\label{sec:background}

\subsection{SWIPE}\label{sec:swipebackground}

\begin{figure}
  \centering
  \includegraphics[alt={SWIPE kernel illustration},width=0.95\linewidth]{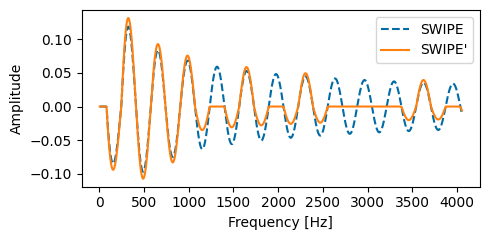}
  \caption{SWIPE and SWIPE' kernels corresponding to a pitch candidate at 330 Hz. The SWIPE kernel contains peaks at all integer harmonics of the candidate frequency. The SWIPE' kernel is obtained by removing the peaks at non-prime harmonics.}
  \label{fig:swipe}
\end{figure}

The \emph{Sawtooth Waveform Inspired Pitch Estimator} (SWIPE) \cite{camachoSawtoothWaveformInspired2008} estimates pitch by identifying the fundamental frequency of a sawtooth waveform whose spectrum best matches that of the input signal. To achieve this, it constructs spectral kernels for a number of discrete pitch candidates, and assigns a score to each pitch candidate by measuring the similarity between its associated kernel and the spectrum of the input signal.

\subsubsection{Score Calculation}

More formally, consider a (windowed) audio signal $x[n]$ of length $N$ and its discrete Fourier Transform (DFT) $X[k]$, which may be truncated to the $K = \lfloor N/2\rfloor + 1$ bins corresponding to non-negative frequencies if $x$ is real-valued. Let now $C = \{f_1, f_2, \ldots f_{|C|}\}$ be a set of $|C|$ pitch candidates. Then $S_c[k]$ is the spectral kernel associated with the pitch candidate $f_c$, and we can compute its SWIPE score $Z(f_c) : C \rightarrow \mathbb [\hspace{-.1cm} -\hspace{-.1cm} 1, 1]$ as the normalized inner product between $S_c$ and $X$:

\begin{equation}\label{eq:swipe}
    Z(f_c) = \frac{\sum_{k=0}^{K-1} S_c[k]\cdot \left|X[k]\right|^{1/2}}{\left(\sum_{k=0}^{K-1}|X[k]|\right)^{1/2}}
\end{equation}

The pitch estimate is then given by the $f_c$ that maximizes $Z(f_c)$ and may optionally be further refined by e.g. parabolic interpolation of the local maximum.

While \eqnref{eq:swipe} computes the inner product over all bins of the DFT for notational simplicity, the original SWIPE paper suggests resampling the spectrum to the Equivalent Rectangular Bandwidth (ERB) \cite{GLASBERG1990103} scale for speech data, or to the mel scale for musical instruments.

\subsubsection{Kernel Design}

The kernel $S_c$ is designed to maximize the inner product with $X$ if $x$ is a signal with fundamental frequency $f_0=f_c$. To achieve that, it contains cosine lobes of width $f_c/2$ at all integer harmonics of $f_c$, decaying in magnitude to mimic the spectrum of a sawtooth wave. As the authors of SWIPE lay out, this corresponds precisely to the square root of the main lobes of a Hann-windowed sawtooth wave if the size of the analysis window is exactly $T=8/f_c$. The kernel further contains negative-valued valleys at $\frac{1}{2}f_c, \frac{3}{2}f_c, \ldots$, i.e.\ at the midpoint between each harmonic peak.

Since all harmonics of a signal with a true fundamental frequency of $f$ also contribute to the scores of the pitch candidates at $f/2, f/3, \ldots$, a common variant of the SWIPE algorithm removes the non-prime harmonics (except for the first one) of all kernels to reduce the problem of octave errors. This is known as SWIPE', but effective and widespread enough that it is often simply referred to as SWIPE, for example in the Speech Processing Toolkit (SPTK) \cite{sp-nitech2023sptk} implementation, which is based directly on the MATLAB code published along with SWIPE. We take the same approach in this paper and will generally assume that the scores $Z$ are calculated using SWIPE' kernels. An example of such a kernel is illustrated in Figure~\ref{fig:swipe}.

\subsection{Supervised Neural Pitch Estimation}

The established way of using neural networks for pitch estimation is to interpret an audio signal $x[n]$ as a vector $\mathbf{x}$. A network $f_\theta$ then maps $\mathbf{x}$ to a vector $\mathbf{y} \in[0, 1]^{|C|}$, where each entry $y_c$ represents the probability that a corresponding pitch candidate $f_c$ is the pitch of $\mathbf{x}$. In supervised training, this is treated as a multi-class classification problem, calculating the loss using the cross-entropy to the ground truth pitch. The ground truth distribution may be smoothed using Gaussian blurring to aid training \cite{kimCREPEConvolutionalRepresentation2018}. Voicing confidence can be deduced from the entropy of the predicted probability distribution \cite{morrisonCrossdomainNeuralPitch2023a}.

\subsection{Self-Supervised Neural Pitch Estimation}

\emph{Pitch Estimation with Self-Supervised Transposition-Equivariant Objective} (PESTO) \cite{riouPESTOPitchEstimation2023} is a state-of-the-art architecture for self-supervised training of neural network pitch estimators, where natural symmetries of the input are exploited to learn a translation-equivariant representation, instead of providing ground truth pitches to the model.

\subsubsection{Training Setup}

During training, the model learns to optimize a combination of three losses:

An \textbf{equivariance loss} enforces that pitch-shifted versions of an input should result in output distributions that are transpositions of the original input. Given an input $\mathbf{x}$ and its pitch-shifted version $\mathbf{x}^{(k)}$ (shifted by $k$ semitones), their respective outputs $\mathbf{y}$ and $\mathbf{y}^{(k)}$ should satisfy $\phi(\mathbf{y}^{(k)}) = \alpha^k\phi(\mathbf{y})$, where $\phi$ is a deterministic linear mapping:
\begin{equation}
\begin{array}{ccccc}
     \phi &: &\mathbb{R}^{|C|} & \rightarrow & \mathbb{R}  \\
     & & \mathbf{y}& \mapsto &(\alpha, \alpha^2, \ldots, \alpha^{|C|})\mathbf{y}
\end{array} 
\end{equation}
and $\alpha$ is a hyperparameter. 

A \textbf{regularization loss} further ensures that the network's outputs for pitch-shifted inputs maintain the expected transposition relationship. For a pair of outputs $\mathbf{y}$ and $\mathbf{y}^{(k)}$, the shifted cross-entropy loss
\begin{equation}
    \mathcal{L}_{\text{SCE}}(\mathbf{y}, \mathbf{y}^{(k)}, k) = \sum_{i=0}^{|C|-1}y_i\log\left(y_{i+k}^{(k)}\right)
\end{equation}
measures how well $\mathbf{y}^{(k)}$ matches the $k$-semitone shift of $\mathbf{y}$.

Finally, an \textbf{invariance loss} encourages the mapping $f_\theta$ to be invariant to the timbre of the signal. During training, PESTO draws random transforms $t$ from a set of pitch-preserving data augmentations $\mathcal{T}$. Given $\tilde{\mathbf{x}} = t(\mathbf{x})$, the invariance loss is then expressed as the cross-entropy between $\mathbf{y} = f_\theta(\mathbf{x})$ and $\tilde{\mathbf{y}} = f_\theta({\mathbf{\tilde{x}}})$.

\subsubsection{Model Architecture}

The PESTO architecture uses the constant-Q transform (CQT) of an audio frame as its input, where the bins of the CQT exactly correspond to the pitch candidates. The transforms $\mathcal{T}$ take the form of adding random noise and gain to the CQT frames. A CNN processes the frame, and its flattened output is fed to a final linear layer followed by a softmax layer which produces a probability distribution. Importantly, the final linear layer uses a Toeplitz matrix as its weight matrix to preserve the transposition equivariance of the CNN.

\section{Methods}\label{sec:methods}

The core insight of this work is that SWIPE scores encode rich pitch information and are thus well suited as an audio frontend for neural pitch estimation in both supervised and self-supervised settings. This section first covers our implementation of SWIPE in detail, and then describes how we adapt supervised and self-supervised neural pitch estimators to work with SWIPE scores.

\subsection{SWIPE Implementation}\label{sec:swipeimpl}

\begin{table*}
    \centering
    \begin{tabular}{c|ccccc}
       & Reported in \cite{gfellerSPICESelfSupervisedPitch2020, engelSelfSupervisedPitchDetection2020} & SPTK ($2 \text{ kHz}$) & SPTK ($8 \text{ kHz}$) & Ours (ERB) & Ours (mel) \\\hline
       MIR-1K & 86.6\% & 96.5\% & 68.2\% & 95.7\% & 96.2\%  \\
       MDB & 90.7\% & 94.1\% & 61.4\% & 94.0\% & 96.1\%
         
    \end{tabular}
    \caption{Raw Pitch Accuracy obtained by SWIPE implementations on the MIR-1K and MDB-stem-synth datasets, compared to baseline values previously reported in self-supervised pitch estimation papers. For SPTK, we report different upper limits for the search range. For our implementation, the search range is constant, but the frequency sampling scale changes.}
    \label{tab:swipecomp}
\end{table*}

We calculate the SWIPE scores by sampling the spectrum at 1024 frequencies, which are linearly spaced on the mel scale over a range from $0.25\cdot f_{\min}$ to $1.25\cdot f_{\max}$. We used the Slaney-style mel scale, which is linear up to 1 kHz and logarithmic above, as implemented in the librosa toolkit \cite{mcfee2015librosa}. For each of the various window sizes, the spectrum is calculated with the same FFT resolution (zero-padding the input as needed) and evaluated at the sampling frequencies using linear interpolation. We arrange the pitch candidates to match the CQT resolution used in PESTO:  logarithmically spaced over a range of $f_{\min} = 27.5\text{ Hz}$ to $f_{\max} = 8055\text{ Hz}$, using a resolution of 3 bins per semitone, for a total of 295 bins. 

Although many papers on neural pitch estimation compare their work to SWIPE as a baseline, they generally do not cite the implementation they used or report the parameters they chose. To make sure that our implementation does not significantly underperform, we compare it to the most popular open-source implementation of SWIPE, which is contained in the Speech Processing Toolkit (SPTK) \cite{sp-nitech2023sptk}. It uses 8 pitch bins per semitone by default, samples the input frequency spectrum according to the ERB \cite{GLASBERG1990103} scale, and refines the estimate using parabolic interpolation. 

Table~\ref{tab:swipecomp} contains the Raw Pitch Accuracy (RPA) achieved by the SPTK and our implementation on the MDB-stem-synth and MIR-1K datasets and compares it to previously reported baseline values. The metrics and datasets are described in more detail in Section~\ref{sec:setup}. The accuracy of the SPTK implementation seems to significantly deteriorate for large search ranges. We report the values for upper limits of $2 \text{ kHz}$ and $8 \text{ kHz}$, where the lower limit is $30\text{ Hz}$ for both. We set the score threshold which pitch candidates need to exceed to be considered to 0. 

Our implementation appears to be a lot more robust to its large search range (27.5--8055 Hz). Switching from ERB to mel sampling results in a notable accuracy gain on MDB-stem-synth, which contains more varied timbres. This matches the results of the original SWIPE paper. 

Both the SPTK and our own implementation can perform much more accurately than the values that were previously reported as baselines in the neural pitch estimation literature suggest, to the extent that SWIPE outperforms even state-of-the-art self-supervised pitch detection models (see Section~\ref{sec:selfsupresults}).

\subsection{Supervised Neural Pitch Estimation}\label{sec:sup}

We experiment with using both SWIPE scores and the CQT as an audio frontend in a supervised training context.

We feed the input into a CNN with 6 1D-convolutional layers, applying layer normalization and a leaky ReLU non-linearity with slope 0.3 between each layer. Zero-padding is applied to the input in each layer to preserve the input dimension. After flattening, the output of the final layer is reduced to the dimensionality of the pitch bins and fed into a Softmax layer to obtain a probability distribution. We find that using a dense linear layer to perform the dimensionality reduction strongly degraded generalization in this setup, and instead also employ a Toeplitz layer in the supervised model.

\subsection{Self-Supervised Neural Pitch Estimation}\label{sec:selfsup}

The original PESTO architecture is already well suited to work with SWIPE scores, which can be directly substituted for the CQT bins without violating the assumptions on translation equivariance. Since SWIPE scores encode periodicity information much more explicitly than CQT frames, we expect the encoder network to achieve similar performance with fewer parameters. We test this hypothesis by training a PESTO-style encoder with a drastically reduced parameter count, consisting only of the final Toeplitz fully-connected layer -- a convolutional layer with a single filter of size 647 -- and softmax normalization. In this very simple architecture, the Toeplitz layer can be seen as essentially learning a reweighting of the SWIPE scores, and to refine the location of the peak score if the output resolution is larger than the input resolution.

Initial experiments indicated that applying random data augmentation to the SWIPE scores only resulted in degraded performance compared to the baseline DSP algorithm. We additionally augment the audio frame in the time domain by adding random noise and applying a finite impulse response (FIR) filter with a randomized amplitude response. This results in an increased computational cost for training, since the SWIPE scores need to be recalculated at every training step, but does not affect the computational cost of inference once training has finished.

\section{Experimental Setup}\label{sec:setup}

\subsection{Datasets}

Our experiments use three $f_0$-annotated datasets that are commonly used for training and benchmarking pitch detectors:

\textbf{MDB-stem-synth} \cite{salamon2017analysis} contains 230 solo tracks (418 minutes total) of instrument sounds and vocals. The audio is re-synthesized from its $f_0$ annotations, which means that the $f_0$ annotations are perfect. It is annotated with a hop size of 2.9 ms.

\textbf{PTDB-TUG} \cite{pirker2011pitch} contains 4720 audio and laryngograph recordings (576 minutes total) of 20 English speakers reading sentences. It is annotated with a hop size of 10 ms.

\textbf{MIR-1K} \cite{5153305} contains 1000 short recordings (133 minutes total) of Chinese karaoke performances. It is annotated with a hop size of 20 ms.

\subsection{Baselines}

We compare all results to two DSP-based pitch detection baselines: PYIN \cite{mauch2014pyin} and SWIPE. We do not perform Viterbi decoding or any sort of peak refinement, simply selecting the pitch candidate with the highest score. PYIN scores are based on autocorrelation and so their natural resolution is expressed in integer samples. We resample the scores to the same pitch candidate resolution as used for SWIPE using linear interpolation. 

As a baseline for supervised training, we choose FCNF0++ \cite{morrisonCrossdomainNeuralPitch2023a}, which to the best of our knowledge is the currently best-performing supervised monophonic neural pitch detector that operates on a frame-by-frame basis, rather than processing the entire audio signal at once.

In the self-supervised setting, we compare our results against the original PESTO architecture, which is the current state of the art in self-supervised monophonic neural pitch estimation.

\subsection{Evaluation Metrics}

We use the {mir\_eval} package \cite{raffel2014mir_eval} to report the following metrics: 

\textbf{Raw Pitch Accuracy} (RPA), the percentage of voiced frames for which the model predicted a pitch within 50 cents of the ground truth.

\textbf{F-Score}, measuring the accuracy of the binary voiced/unvoiced decision.

\textbf{Overall Accuracy} (OA), the percentage of all frames (voiced and unvoiced) for which a correct voicing decision was made, and for which the model predicted a pitch within 50 cents of the ground truth if the frame is voiced.

\section{Results and Discussion}\label{sec:results}

We report separate experimental results for the supervised and self-supervised approaches, in each case closely replicating the training setup of the baselines (FCNF0++ and PESTO, respectively) to assess the impact of using SWIPE scores as an audio frontend.

\subsection{Supervised Models}

We refer to the two proposed supervised models (see Section~\ref{sec:sup}) as \textbf{CQT-sup} and \textbf{SWIPE-sup}. We train our models as well as the FCNF0++ baseline on {MDB-stem-synth} and {PTDB-TUG} at the same time. While networks that take CQT or SWIPE scores as input are sample-rate agnostic, FCFN0++ operates on the raw audio waveform and was designed to work with a sampling rate of $8\text{ kHz}$, so we resample its input accordingly.

In the interest of a direct comparison, we use the 70-15-15 split into training, validation and testing partitions that was published in \cite{morrisonCrossdomainNeuralPitch2023a}. The performance of the trained models is measured by calculating RPA, F-Score and OA on the testing set. To better measure generalization performance on unseen data, we additionally evaluate the trained models on the full {MIR-1K} dataset, which is not used in training.


\begin{table*}[h]
\centering
\begin{tabular}{l|cc|ccc|ccc}
\multirow{2}{*}{Method} & \multicolumn{1}{l}{\multirow{2}{*}{\# Params}} & \multicolumn{1}{l|}{\multirow{2}{*}{Window Size {[}ms{]}}} & \multicolumn{3}{c|}{Test Partition}                 & \multicolumn{3}{c}{{MIR-1K}}                          \\
                        & \multicolumn{1}{l}{}                           & \multicolumn{1}{l|}{}                                      & RPA             & F-Score         & OA              & RPA             & F-Score         & OA              \\ \hline
PYIN                    & -                                              & 145                                                        & 89.4\%          & -               & -               & 95.4\%          & -               & -               \\
SWIPE                   & -                                              & 327                                                        & 93.2\%          & -               & -               & \textbf{96.2\%}          & -               & -               \\ \hline
CQT-sup (ours)          & 0.9M                                           & 1871                                                       & 98.1\%          & \textbf{98.7\%} & \textbf{99.2\%} & 92.2\%          & \textbf{91.2\%} & 87.1\%          \\
SWIPE-sup (ours)        & 0.9M                                           & 327                                                        & 97.9\%          & 98.1\%          & 98.6\%          & 93.5\% & 90.0\%          & \textbf{87.8\%} \\ \hline
FCNF0++ \cite{morrisonCrossdomainNeuralPitch2023a}                & 6.6M                                           & 128                                                        & \textbf{98.3\%} & 98.2\%          & 98.7\%          & 91.0\%          & 87.8\%          & 86.0\%          \\
\end{tabular}
\caption{Evaluation results of the supervised models as measured by Raw Pitch Accuracy (RPA), the F-Score for the voiced/unvoiced decision, and Overall Accuracy (OA). We also report the sizes of the neural networks and the maximum window size required by the estimator. Models are evaluated using the combined test partition of {MDB-stem-synth} and {PTDB-TUG} published in
\cite{morrisonCrossdomainNeuralPitch2023a}, as well as on the entire {MIR-1K} dataset, which was not used in training. }
\label{tab:supresults}
\end{table*}

We train the models for 500,000 steps, using a batch size of 256 and the Adam optimizer \cite{kingmaAdamMethodStochastic2015} with an initial learning rate of 0.0002. Table~\ref{tab:supresults} shows the results of the evaluation. Overall, both {CQT-sup} and {SWIPE-sup} seem competitive with FCNF0++, but not clearly superior. They are able to almost match the in-domain RPA of FCNF0++, and outperform it in terms of voiced/unvoiced accuracy and generalization. The two proposed models use fewer trainable parameters than FCNF0++, but require a larger context window. 

The CQT input features seem to be particularly well suited for making voiced/unvoiced decisions, with the {CQT-sup} model attaining the highest F-Score on both the test set and on {MIR-1K}. 

All three models struggle with generalization, staying well behind the DSP baselines on MIR-1K. The best generalization behavior is shown by {SWIPE-sup}, even though it was the least accurate model on the test set.

\subsection{Self-Supervised Models}\label{sec:selfsupresults}
We refer to the three modified PESTO models (see Section~\ref{sec:selfsup}) as \textbf{CQT-tiny}, \textbf{SWIPE-full}, and \textbf{SWIPE-tiny}, where ``tiny'' refers to the Toeplitz-only encoder and ``full'' to the original PESTO encoder architecture with a multi-layer CNN. We train the models on the whole of MIR-1K and measure their performance on MDB-stem-synth, and vice versa. The models are trained for 50 epochs using a batch size of 256 and the Adam optimizer with an initial learning rate of 0.0001.

\begin{table*}[h]
\centering
\begin{tabular}{lcccc}
                                                                                               & \multicolumn{1}{l}{}   & \multicolumn{1}{l}{}    & \multicolumn{2}{c}{Raw Pitch Accuracy}    \\
Method                                                                                         & \# params              & Trained on              & {MIR-1K} & {MDB-stem-synth} \\ \hline
PYIN                                                                                           & -                      & -                       & 95.4\%          & 91.6\%                  \\
SWIPE                                                                                          & -                      & -                       & 96.2\%          & 96.1\%                  \\ \hline
\multirow{2}{*}{\begin{tabular}[c]{@{}l@{}}PESTO\\ (baseline from \cite{riouPESTOPitchEstimation2023})\end{tabular}} & \multirow{2}{*}{28.9k} & {MIR-1K}         & 96.1\%          & \textbf{94.6\%}                  \\
                                                                                               &                        & {MDB-stem-synth} & 93.5\%          & 95.5\%                  \\ \hline
\multirow{2}{*}{CQT-tiny}                                                                & \multirow{2}{*}{647}   & {MIR-1K}         & 95.6\%          & 78.8\%                  \\
                                                                                               &                        & {MDB-stem-synth} &        91.7\%         & 95.5\%                        \\ \hline
\multirow{2}{*}{SWIPE-full}                                                              & \multirow{2}{*}{28.2k} & {MIR-1K}         & \textbf{97.0\%} & 89.7\%                  \\
                                                                                               &                        & {MDB-stem-synth} & 96.1\%          & 96.4\%                  \\ \hline
\multirow{2}{*}{SWIPE-tiny}                                                              & \multirow{2}{*}{647}   & {MIR-1K}         & 96.6\%          & 90.1\%                  \\

                                                                                               &                        & {MDB-stem-synth} & \textbf{96.4\%}          & \textbf{96.5\%}         \\ 
\end{tabular}
  \caption{Evaluation results of the self-supervised models. For both datasets, we highlight the best result achieved when training and evaluating on the same dataset (no explicit pitch information is provided to the model during training), and when training on one dataset and evaluating on the other. The performance of PYIN and SWIPE is given for comparison.}
  \label{tab:sslresults}
\end{table*}

The results of the evaluation are given in Table~\ref{tab:sslresults}. The baseline SWIPE implementation outperforms PESTO on both datasets, regardless of which dataset the model was trained on. \textbf{This means that the original SWIPE algorithm outperforms all work on self-supervised monophonic pitch detection published to date.}

When using CQT frames as input, reducing the encoder network to just the final Toeplitz layer noticeably degrades performance, especially in the across-dataset evaluation. However, it is worth noting that CQT-tiny still achieves relatively good same-dataset accuracy. Since no explicit pitch information is given to the model during training, this is a strong indicator for the usefulness of the transposition-equivariant training structure that PESTO introduced.

The highest accuracy on the same-dataset evaluation is achieved by the two models that use SWIPE scores as input. Like CQT-tiny however, their performance plummets when trained on MIR-1K and evaluated on MDB-stem-synth. MDB-stem-synth covers a larger pitch range than MIR-1K and contains more varied timbres, making generalization challenging. The original PESTO is the only model that is able to make this jump reasonably well.

In the reverse direction however, SWIPE-tiny achieves the highest RPA out of the four models when trained on MDB-stem-synth and evaluated on MIR-1K, as well as the best same-dataset RPA for MDB-stem-synth. Adding the Toeplitz layer on top of the SWIPE scores improves their performance, but the additional network layers in SWIPE-full do not bring further accuracy gains, seemingly hindering performance instead.

\begin{figure}[h]
    \centering
    \includegraphics[width=\linewidth]{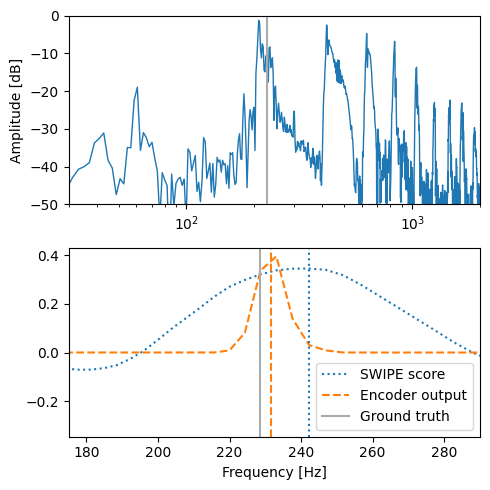}
    \caption{\textbf{Top:} The spectrum of a frame of audio from MIR-1K. The solid vertical line marks the ground truth pitch. \textbf{Bottom:} The SWIPE scores for the frame, before (dotted) and after (dashed) they were transformed by the SWIPE-tiny encoder trained on MDB-stem-synth. The vertical lines indicate the pitch estimate obtained from the basic SWIPE algorithm (dotted), the estimate given by SWIPE-tiny (dashed), and the ground truth (solid).}
    \label{fig:exampleframe}
\end{figure}

The Toeplitz-only encoder in SWIPE-tiny seems to learn to refine the peaks of the SWIPE scores, mitigating errors caused by the quantization of the search space or by input spectra that deviate too far from the harmonic ideal. Figure~\ref{fig:exampleframe} illustrates a frame where an estimation error of 90 cents is reduced to 30 cents after feeding the SWIPE scores through the SWIPE-tiny encoder.

\subsection{Latency-Accuracy Tradeoff for SWIPE}\label{sec:adawin}

The frame-based structure of the evaluated models, and especially the lightweight architecture of the self-supervised estimators, lend themselves naturally to use in real-time, streaming applications. In this context, small window sizes are desirable to reduce latency. 

As described in Section~\ref{sec:swipebackground}, the theoretical ideal window size for each pitch candidate $f_c$ is exactly $8 / f_c$. In practice however, it is sufficient to only consider window sizes whose length in samples is a power of two. For a given pitch candidate with an ideal window length $W$, the score is then calculated twice at window lengths $2^{\lfloor\log_2(W)\rfloor}$ and $2^{\lceil\log_2(W)\rceil}$, and linearly interpolated to obtain an approximation to the score at the ideal size. Given a sampling rate of $f_s = 44.1\text{ kHz}$ and a minimum pitch candidate of $f_{\min} = 27.5\text{ Hz}$, the next-longest window with a power-of-two length in samples corresponds to 327 ms, which is already a significant improvement compared to the 1871 ms required by the CQT. 

However, this can be reduced further. The SWIPE-based models offer a straightforward way to reduce both latency and computational cost at the expense of accuracy by simply calculating the scores for lower pitch candidates at shorter window sizes (without interpolation). Crucially, this adjustment can be made flexibly at inference time without  retraining the model. Table~\ref{tab:windowsizes} shows the effect that reducing the window size has on the RPA for two selected models. Note that reducing the window size to a length that is not a power of two is also possible if finer control over the tradeoff is desired.

\begin{table}[]
    \centering
\begin{tabular}{cc|cc}
\multicolumn{2}{c|}{Window Size} & \multicolumn{2}{c}{Raw Pitch Accuracy} \\
{[}Samples{]}     & {[}ms{]}     & SWIPE-tiny               & SWIPE-sup               \\ \hline
16384             & 372          & 96.4\%             & 97.2\%            \\
8192              & 186          & 96.4\%             & 97.1\%            \\
4096              & 93           & 96.2\%             & 96.7\%            \\
2048              & 46           & 85.0\%             & 86.9\%           
\end{tabular}
    \caption{The effect of reducing the maximum window size (for a sampling rate of $44.1\text{ kHz}$) at which SWIPE scores are calculated. RPA on MIR-1K is reported for SWIPE-tiny (trained on MDB-stem-synth) and SWIPE-sup.}
    \label{tab:windowsizes}
\end{table}

\subsection{Robustness to Noise}

In the final experiment, we investigate how robust various trained models are to noisy conditions by adding white noise to the input audio at decreasing signal-to-noise ratios. The results are shown in Table~\ref{tab:noise}. SWIPE-tiny appears to be fairly robust to background noise, especially compared to the base SWIPE algorithm. The performance of the supervised models degrades somewhat quicker than that of the self-supervised ones. This is not too surprising, since invariance to added noise is an explicit training objective for the self-supervised models (see Section~\ref{sec:selfsup}).

\begin{table}[]
    \centering
    \setlength{\tabcolsep}{3.5pt}
\begin{tabular}{l|ccccc}
                 & \multicolumn{5}{c}{Raw Pitch Accuracy (MIR-1K)}                                                        \\
Model            & clean           & 5 dB            & 0 dB            & -5 dB           & -10 dB          \\ \hline
PYIN             & 95.4\%          & 95.3\%          & 95.1\%          & \textbf{93.7\%} & 85.8\%          \\
SWIPE            & 96.2\%          & 93.9\%          & 91.2\%          & 85.6\%          & 75.2\%          \\ \hline
CQT-sup          & 92.2\%          & 91.5\%  & 89.3\%          &   87.3\%      & 82.3\%          \\
SWIPE-sup        & 93.5\% & 91.6\%          & 90.0\%          & 87.1\%          & 72.2\%          \\
SWIPE-tiny & \textbf{96.6}\%          & \textbf{96.0}\%          & \textbf{95.3\%} & 93.4\%          & \textbf{88.5\%} \\ \hline
PESTO   & 94.6\%          & 93.3\%          & 92.9\%          & 90.1\%          & 81.7\%          \\
FCNF0++          & 91.0\%          & 90.3\%          & 89.0\%          & 83.5\%          & 81.0\%         
\end{tabular}
    \caption{The effect that adding white noise at various signal-to-noise ratios to the input audio has on the raw pitch accuracy on the MIR-1K dataset for various models.}
    \label{tab:noise}
\end{table}

\section{Conclusion}

We investigated the potential of combining the SWIPE algorithm with neural pitch estimation. We adapted established supervised and self-supervised training techniques to use SWIPE scores as an audio frontend and obtained accurate, efficient, robust and flexible pitch estimators. 

We demonstrated that the potential of SWIPE has been significantly underestimated in the literature despite being commonly used as a baseline. The algorithm in its original form outperforms state-of-the-art self-supervised neural pitch estimators.

In the future, we plan to explore whether the performance of the pitch estimators can be further improved by hybrid training schemes that make simultaneous use of labeled data and self-supervised training objectives. We would also like to investigate whether certain parameters of SWIPE, such as the frequency envelope or the weights of individual harmonics, can be directly learned from data.

\clearpage

\section{Acknowledgments}
This work was supported by UK Research and Innovation [grant number EP/S022694/1]. The authors would like to thank the anonymous reviewers for their valuable feedback which significantly improved this paper.

\bibliography{pitch}

\end{document}